\documentclass[longbibliography,superscriptaddress, amsmath, aps, prl,  numerical, twocolumn]{revtex4-1} 

\usepackage[T1]{fontenc}
\usepackage{latexsym}
\usepackage{amssymb}
\usepackage{amsfonts}
\usepackage{amsmath}
\usepackage{amsthm}
\usepackage{multirow}
\usepackage{longtable}
\usepackage{mathtools}
\usepackage{upgreek}
\usepackage{braket}
\usepackage{siunitx}
\usepackage{xcolor}
\usepackage{hyperref}

\newcommand{\dmscr}{\delta \left\langle r_\mathrm{c}^2 \right\rangle}
\newcommand{\mscr}{\left\langle r_\mathrm{c}^2 \right\rangle}

\def\orcid#1{\kern .08em\href{https://orcid.org/#1}{\includegraphics[keepaspectratio,width=0.7em]{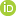}}}

\makeatletter
\def\@bibdataout@aps{%
\immediate\write\@bibdataout{%
@CONTROL{%
apsrev41Control%
\longbibliography@sw{%
    ,author="08",editor="1",pages="0",title="0",year="1"%
    }{%
    ,author="08",editor="1",pages="0",title="",year="1"%
    }%
  }%
}%
\if@filesw \immediate \write \@auxout {\string \citation {apsrev41Control}}\fi 
}
\makeatother

\begin{document}

\title{Nuclear Charge Radii of the Nickel Isotopes $^{58-68,70}$Ni}

\author{S. Malbrunot-Ettenauer}
\email{stephan.ettenauer@cern.ch}
\affiliation{Experimental Physics Department, CERN, CH-1211 Geneva 23, Switzerland}
\author{S. Kaufmann}
\email{present address:
Institut f\"ur Kernchemie, Johannes Gutenberg-Universit\"at Mainz, D-55128 Mainz, Germany}
\affiliation{Institut f\"ur Kernphysik, Technische Universit\"at Darmstadt, D-64289 Darmstadt, Germany}
\author{S.~Bacca\orcid{0000-0002-9189-9458}}
\affiliation{Institut f\"ur Kernphysik and PRISMA$^+$ Cluster of Excellence, Johannes Gutenberg-Universit\"at Mainz, D-55128 Mainz,
  Germany}
\affiliation{Helmholtz-Institut Mainz, GSI Helmholtzzentrum f\"ur Schwerionenforschung GmbH, D-64291 Darmstadt, Germany}
\author{C. Barbieri \orcid{0000-0001-8658-6927}}
\affiliation{Department of Physics, University of Surrey, Guildford, GU2 7XH, United Kingdom}
\affiliation{Dipartimento di Fisica, Universit\`a degli Studi di Milano, Via Celoria 16, 20133 Milano, Italy}
\affiliation{INFN, Sezione di Milano, Via Celoria 16, 20133 Milano, Italy}
\author{J.~Billowes}
\affiliation{School of Physics and Astronomy, The University of Manchester, Manchester, M13 9PL, United Kingdom}
\author{M.~L.~Bissell}
\affiliation{School of Physics and Astronomy, The University of Manchester, Manchester, M13 9PL, United Kingdom}
\author{K.~Blaum~\orcid{0000-0003-4468-9316}}
\affiliation{Max-Planck-Institut f\"ur Kernphysik, D-69117 Heidelberg, Germany}
\author{B.~Cheal}
\affiliation{Oliver Lodge Laboratory, Oxford Street, University of Liverpool, Liverpool, L69 7ZE, United Kingdom}
\author{T.~Duguet}
\affiliation{IRFU, CEA, Universit\'e Paris-Saclay, 91191 Gif-sur-Yvette, France} 
\affiliation{KU Leuven, Instituut voor Kern- en Stralingsfysica, B-3001 Leuven, Belgium}
\author{R.~F.~Garcia~Ruiz}
\email{present address:
Massachusetts Institute of Technology, Cambridge, MA, USA}
\affiliation{School of Physics and Astronomy, The University of Manchester, Manchester, M13 9PL, United Kingdom}
\affiliation{Experimental Physics Department, CERN, CH-1211 Geneva 23, Switzerland}
\author{W.~Gins}
\email{present address: Department of Physics, University of Jyv\"askyl\"a, P.O. Box 35 (YFL), FI-40014 Jyv\"askyl\"a, Finland}
\affiliation{KU Leuven, Instituut voor Kern- en Stralingsfysica, B-3001 Leuven, Belgium}
\author{C.~Gorges}
\affiliation{Institut f\"ur Kernphysik, Technische Universit\"at Darmstadt, D-64289 Darmstadt, Germany}
\author{G.~Hagen ~\orcid{0000-0001-6019-1687}}
\affiliation{Physics Division, Oak Ridge National Laboratory, Oak
  Ridge, Tennessee 37831, USA}
\affiliation{Department of Physics and Astronomy, University of Tennessee, Knoxville, TN 37996, USA}
\author{H.~Heylen}
\affiliation{Max-Planck-Institut f\"ur Kernphysik, D-69117 Heidelberg, Germany}
\affiliation{Experimental Physics Department, CERN, CH-1211 Geneva 23, Switzerland}
\author{J.~D.~Holt \orcid{0000-0003-4833-7959}}%
\affiliation{TRIUMF 4004 Wesbrook Mall, Vancouver BC V6T 2A3, Canada}%
\affiliation{Department of Physics, McGill University, Montr\'eal, QC H3A 2T8, Canada}%
\author{G.~R.~Jansen~\orcid{0000-0003-3558-0968} }
\affiliation{Physics Division, Oak Ridge National Laboratory, Oak
  Ridge, Tennessee 37831, USA}
\affiliation{National Center for Computational Sciences, Oak Ridge National Laboratory, Oak Ridge, TN 37831, USA}
\author{A.~Kanellakopoulos}
\email{Present address: HEPIA Geneva, HES-SO, 1202 Geneva, Switzerland}
\affiliation{KU Leuven, Instituut voor Kern- en Stralingsfysica, B-3001 Leuven, Belgium}
\author{M.~Kortelainen~\orcid{0000-0001-6244-764X}}
\affiliation{Department of Physics, University of Jyv\"askyl\"a, P.O. Box 35 (YFL), FI-40014 University of Jyv\"askyl\"a, Finland
}
\author{T. Miyagi}%
\affiliation{TRIUMF 4004 Wesbrook Mall, Vancouver BC V6T 2A3, Canada}%
\author{P. Navr\'atil~\orcid{0000-0002-6535-2141}}
\affiliation{TRIUMF 4004 Wesbrook Mall, Vancouver BC V6T 2A3, Canada}
 \author{W.~Nazarewicz\orcid{0000-0002-8084-7425}}
 \affiliation{Department of Physics and Astronomy and FRIB Laboratory, Michigan State University, East Lansing, Michigan 48824, USA}
\author{R.~Neugart}
\affiliation{Max-Planck-Institut f\"ur Kernphysik, D-69117 Heidelberg, Germany}
\affiliation{Institut f\"ur Kernchemie, Johannes Gutenberg-Universit\"at Mainz, D-55128 Mainz, Germany}
\author{G.~Neyens}
\affiliation{Experimental Physics Department, CERN, CH-1211 Geneva 23, Switzerland}
\affiliation{KU Leuven, Instituut voor Kern- en Stralingsfysica, B-3001 Leuven, Belgium}
\author{W. N\"ortersh\"auser ~\orcid{0000-0001-7432-3687}}
\email{wnoertershaeuser@ikp.tu-darmstadt.de}
\affiliation{Institut f\"ur Kernphysik, Technische Universit\"at Darmstadt, D-64289 Darmstadt, Germany}
\author{S.~J.~Novario}
\affiliation{Department of Physics and Astronomy, University of Tennessee, Knoxville, TN 37996, USA}
\affiliation{Physics Division, Oak Ridge National Laboratory, Oak
  Ridge, Tennessee 37831, USA}
\author{T.~Papenbrock~\orcid{0000-0001-8733-2849}}
\affiliation{Department of Physics and Astronomy, University of Tennessee, Knoxville, TN 37996, USA}
\affiliation{Physics Division, Oak Ridge National Laboratory, Oak
  Ridge, Tennessee 37831, USA}
\author{T.~Ratajczyk}
\affiliation{Institut f\"ur Kernphysik, Technische Universit\"at Darmstadt, D-64289 Darmstadt, Germany}
\author{P.-G.~Reinhard}
\affiliation{Institut f\"ur Theoretische Physik II, Universit\"at Erlangen-N\"urnberg, 91058 Erlangen, Germany}
\author{L.~V.~Rodr\'iguez}
\affiliation{Experimental Physics Department, CERN, CH-1211 Geneva 23, Switzerland}
\affiliation{Max-Planck-Institut f\"ur Kernphysik, D-69117 Heidelberg, Germany}
\affiliation{Institut de Physique Nucl\'eaire, CNRS-IN2P3, Universit\'e Paris-Sud, Universit\'e Paris-Saclay, 91406 Orsay, France}
\author{R. S\'anchez~\orcid{0000-0002-4892-4056}}
\affiliation{GSI Helmholtzzentrum f\"ur Schwerionenforschung GmbH, D-64291 Darmstadt, Germany}
\author{S.~Sailer}
\affiliation{Technische Universit\"at M\"unchen, D-80333 M\"unchen, Germany}
\author{A.~Schwenk~\orcid{0000-0001-8027-4076}}
\affiliation{Institut f\"ur Kernphysik, Technische Universit\"at Darmstadt, D-64289 Darmstadt, Germany}
\affiliation{ExtreMe Matter Institute EMMI, GSI Helmholtzzentrum f\"ur Schwerionenforschung GmbH, D-64291 Darmstadt, Germany}
\affiliation{Max-Planck-Institut f\"ur Kernphysik, D-69117 Heidelberg, Germany}
\author{J.~Simonis}
\affiliation{Institut f\"ur Kernphysik and PRISMA$^+$ Cluster of Excellence, Johannes Gutenberg-Universit\"at Mainz, D-55128 Mainz,
  Germany}
\author{V.~Som\`a}
\affiliation{IRFU, CEA, Universit\'e Paris-Saclay, 91191 Gif-sur-Yvette, France}
\author{S.~R.~Stroberg}%
\affiliation{Department of Physics, University of Washington, Seattle, WA 98195, USA}
\author{L.~Wehner}
\affiliation{Institut f\"ur Kernchemie, Johannes Gutenberg-Universit\"at Mainz, D-55128 Mainz, Germany}
\author{C.~Wraith}
\affiliation{Oliver Lodge Laboratory, Oxford Street, University of Liverpool, Liverpool, L69 7ZE, United Kingdom}
\author{L.~Xie}
\affiliation{School of Physics and Astronomy, The University of Manchester, Manchester, M13 9PL, United Kingdom}
\author{Z.~Y.~Xu}
\affiliation{KU Leuven, Instituut voor Kern- en Stralingsfysica, B-3001 Leuven, Belgium}
\author{X.~F.~Yang~\orcid{0000-0002-1633-4000}}
\affiliation{School of Physics and State Key Laboratory of Nuclear Physics and Technology, Peking University, Beijing 100871, China}
\affiliation{KU Leuven, Instituut voor Kern- en Stralingsfysica, B-3001 Leuven, Belgium}
\author{D.~T.~Yordanov}
\affiliation{Institut de Physique Nucl\'eaire, CNRS-IN2P3, Universit\'e Paris-Sud, Universit\'e Paris-Saclay, 91406 Orsay, France}

\begin{abstract} 
Collinear laser spectroscopy is performed on the nickel isotopes $^{58-68,70}$Ni, using a time-resolved photon counting system. From the measured isotope shifts,
nuclear charge radii $R_c$ are extracted and compared to  theoretical results. 
Three \textit{ab initio} approaches all employ, among others, the chiral interaction NNLO$_{\rm sat}$, which allows an assessment of their accuracy. We find agreement with experiment in differential radii $\dmscr$ for all employed \textit{ab initio} methods and interactions, while the absolute radii are consistent with data only for NNLO$_{\rm sat}$. Within nuclear density functional theory,
the Skyrme functional SV-min matches experiment more closely than the Fayans functional Fy($\Delta r$,~HFB).
\end{abstract}
\maketitle
{\it Introduction. ---} The accurate description of rich physics phenomena encountered in atomic nuclei remains a formidable challenge for contemporary nuclear theory. 
The long-term goal of nuclear physics is thus to develop a universal framework to consistently describe atomic nuclei across the entire nuclear chart. 
Research in recent years has led to remarkable advances in nuclear many-body methods~\cite{hagen2015,Morr17Tin,gysbers2019,Hergert20,PhysRevLett.125.182501,Holt19drip,Soma20c} as well as in the development of nuclear forces based on chiral effective field theory (EFT), rooted in symmetries of QCD and based on pion exchanges and short-ranged interactions~\cite{Epelbaum09, Machleidt11, Hammer20}. A significant theoretical effort has been dedicated to the description of electromagnetic properties such as nuclear charge radii $R_c$. Since charge radii can be measured with high accuracy, they serve as robust benchmarks for nuclear theory. Presently, the region of medium- to heavy-mass nuclei constitutes the testing ground for developing the coherent theoretical nuclear framework.
An important element of this endeavor is to connect \textit{ab initio} models to nuclear density functional theory (DFT). In addition to  \textit{ab initio} calculations, well-calibrated energy density functionals, such as the Fayans functional, are capable of
a successful description of nuclear charge radii 
for multiple isotopic chains ranging from potassium ($Z=19$) all the way to tin ($Z=50$) \cite{Koszorus2021,GarciaCaChRadii2016,Miller2019,PhysRevLett.117.252501,deGroote2020,PhysRevLett.121.102501,PhysRevLett.122.192502}.

In this Letter, we report nuclear charge radii of nickel isotopes (Ni, $Z=28$) which, in terms of $R_c$, constitutes the last unexplored "magic" isotopic chain in this mass region. While the charge radius of $^{68}$Ni was reported earlier \cite{kaufmann2020}, we here present additionally the results for $^{59,63,65-67,70}$Ni. The experimental data are compared with two DFT approaches as well as three independent \textit{ab initio} methods based on chiral EFT interactions.

{\it Experiment. ---}
The experiment at ISOLDE/CERN has been described previously in \cite{kaufmann2020}. Details on the general setup can be found in \cite{Neugart.2017}. In brief, Ni isotopes were produced in a uranium carbide target bombarded with  proton pulses of 1.4-GeV energy. 
Ions were formed by resonant laser ionization with RILIS \cite{marsh14} and accelerated in a first and a second beamtime to about $30$\,keV and $40$\,keV, respectively. Different ISOLDE targets were used with the aim to increase production and to suppress isobars, but they behaved comparably. After mass selection in a high-resolution mass separator, the ions were injected into the radio-frequency quadrupole (RFQ) ion beam cooler and buncher ISCOOL \cite{franberg08} where they were accumulated for typically $10-100$\,ms. After extraction as a short ion bunch, 
the ions were transported to the collinear laser spectroscopy beam line COLLAPS, where the beam was superimposed with a co-propagating laser beam. Bunching reduces the otherwise dominant background of scattered laser light compared to a continuous beam \cite{Nieminen.2002}. 
The ion beam energy was determined by the high-voltage applied to ISCOOL, which was recorded by a precision high-voltage divider. In the first beamtime a 30-kV divider was available, while a 50-kV divider was provided by PTB Braunschweig later on. This allowed independent voltage calibrations and the use of a higher beam energy, favorable for laser-spectroscopic resolution. 

Laser spectroscopy on the neutral Ni atoms was performed after neutralization of the ions in a charge-exchange cell \cite{muller83, klose12} filled with potassium vapor. 
A frequency-doubled single-mode cw titanium-sapphire laser stabilized with a high-resolution wavemeter \cite{Verlinde.2020,Konig.2020} was used to excite the $3d^9\,4s\,^{3}\!D_3 \rightarrow 3d^9\,4p\,^{3}\!P_2$ transition at 352.45\,nm. The wavemeter was calibrated regularly with a stabilized helium-neon laser. Fluorescence photons from spontaneous emission were detected by four photomultiplier tubes. 
All isotopes were measured alternating with the reference isotope $^{60}$Ni to compensate for  remaining long-term drifts in ion velocity or laser frequency.

\begin{figure}[t]
	\centering
		\includegraphics[trim=6mm 6mm 7mm 6mm,clip, width=\columnwidth]{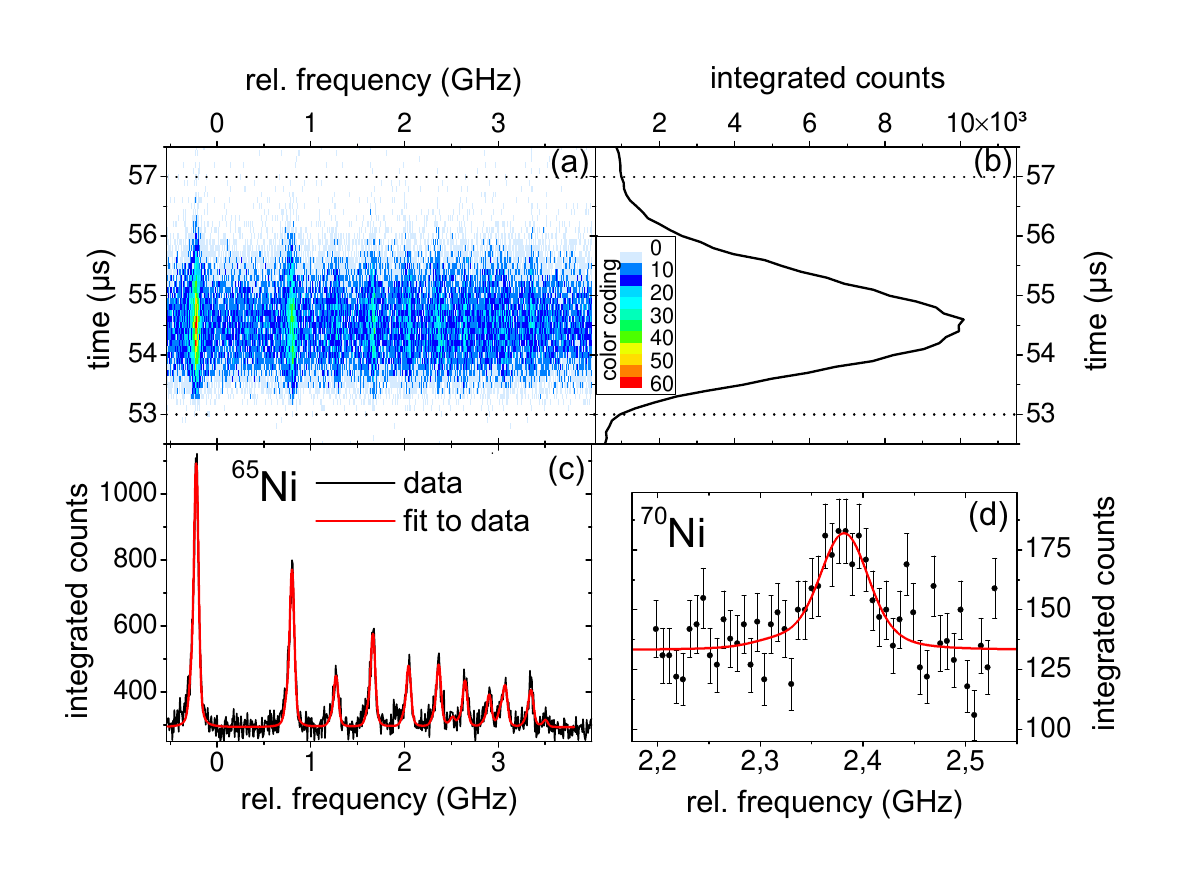}
	\caption{Frequency-time spectrum of a $^{65}$Ni resonance (a), the temporal ion-bunch structure (b) and the laser-spectroscopic resonance spectrum (c). In (d), a resonance of $^{70}$Ni is shown. See text for details.
	}
	\label{fig:f-t-spectrum}
\end{figure}

For the present work, a new data acquisition system called ``TILDA'' \cite{Kaufmann.2015} was employed for the first time
at COLLAPS. It is based on photon tagging with reference to ISCOOL's release trigger \cite{Kanellakopoulos.2020} and relaxes the need for hard-wired gates set during a beamtime. Comparable schemes have previously been employed at other laser experiments with bunched ion beams \cite{PhysRevLett.111.122501,Lochmann.2014,Rossi.2014,GarciaRuiz.2018,DEGROOTE2020437}.  
A typical spectrum recorded with TILDA is shown for $^{65}$Ni in Fig.\,\ref{fig:f-t-spectrum}(a): The $x$-axis represents the laser frequency calculated from the scanning voltage at the charge exchange cell, while the $y$-axis is the time elapsed since the RFQ extraction pulse was recorded. The color represents the number of photons detected within a 100-ns interval during 900 extractions from the RFQ. The time structure of the ion bunch is shown in Fig.\,\ref{fig:f-t-spectrum}(b), where the counts at a specified time are integrated over all frequencies. Similarly, summing all counts at a fixed frequency within the (adjustable) time intervall between 53 and 57\,$\upmu$s reveals the resonance spectrum of the isotope in Fig.\,\ref{fig:f-t-spectrum}(c). 

According to an analysis with ISOLTRAP's multi-reflection time-of-flight mass spectrometer \cite{Wolf.2013}, the beam of the most exotic isotope $^{70}$Ni was dominated by the isobar $^{70}$Ga with a ratio of of $\approx 1:10^4$.  
The large amount of isobaric ions can cause an overfilling of ISCOOL and a corresponding shift in beam-energy due to the ions' space-charge potential, 
which can degrade the accuracy of the spectroscopic measurements. Moreover, non-resonant light emitted by the unwanted ions after collisional excitation or neutralization in the charge exchange cell will reduce the sensitivity for $^{70}$Ni. To suppress $^{70}$Ga, we took advantage of the different target-release properties of the two elements: the beam gate at ISOLDE, allowing the ions to be transported to the experiments, stayed closed during the first 2\,s after the proton impact. Then, most of the more volatile $^{70}$Ga had been released from the target while the remaining fraction of $^{70}$Ni ($T_{1/2} = 6$\,s) was accumulated for 1.2\,s in the RFQ and then sent as a single bunch to COLLAPS, before the next proton pulse arrived. A $^{70}$Ni resonance is shown in Fig.\,\ref{fig:f-t-spectrum}(d).


{\it Analysis. ---}
Isotope shifts $\delta \nu^{60,A} =  \nu^{A} - \nu^{60}$ for all isotopes were calculated from their respective center frequency $\nu^{A}$ with respect to the center frequency $\nu^{60}$ of the reference isotope $^{60}$Ni. 
Both beamtimes were analyzed individually and a linear displacement in their isotope shifts was corrected by introducing a correction to the main acceleration voltage within the uncertainty of the corresponding voltage dividers.
The main acceleration voltage of 30\,kV ($1^\mathrm{st}$\,beamtime) was reduced by 3.5\,V and the 40\,kV ($2^\mathrm{nd}$\,beamtime) was increased by 2.5\,V in the analysis. 
A still remaining scatter in the isotope shifts of individual isotopes as obtained in the two beamtimes  could not be explained by their statistical 
uncertainties. However, this variation was not systematic and could not be traced back to definite reasons individually. Therefore, an additional statistical uncertainty was added to all isotopes, such that the scatter appeared statistically reasonable, i.e., the $\chi_{\mathrm{red}}^2$ calculated from the deviations between the final isotope shifts of the two beamtimes and their average was reduced to 1.
Results are listed in Table \ref{tab:shifts}. 
The changes in mean-square nuclear charge radii $\dmscr^{60,A}\equiv\left\langle r_\mathrm{c}^2 \right\rangle^A-\left\langle r_\mathrm{c}^2 \right\rangle^{60}$ are obtained using the field-shift factor $F=-783(94)$\,MHz/fm$^2$ and the mass-shift factor $M_{\alpha=396}=950(5)$\,GHz\,u, as explained in \cite{kaufmann2020}. These values are in excellent agreement with independent measurements reported in \cite{koenig2021b}. Negligible deviations from our values in \cite{kaufmann2020} arise from a correction in the analysis code but lead only to insignificant changes of $\dmscr$ values. The  uncertainties of  $\dmscr^{60,A}$ are dominated by the correlated error based on the uncertainty of $F$. The absolute charge radii $R_c \equiv \mscr^{1/2}$ are obtained from $\dmscr^{60,A}$ by utilising $R_c(^{60}\mathrm{Ni})=3.806(2)$~fm \cite{Fricke2004}.
\begin{table}[b]
	\centering
		\caption{Measured isotope shifts $\delta \nu^{60,A}$ relative to $^{60}$Ni with statistical uncertainties in parentheses and systematic uncertainties in square brackets. Values for the stable isotopes and $^{68}$Ni are those from \cite{kaufmann2020}. The statistical uncertainty includes variations between the two beamtimes that are partially of systematic but uncorrelated origin and change statistically from isotope to isotope, while the systematic uncertainty is restricted to the correlated uncertainty caused by the high-voltage measurement. The extracted change in mean-square charge radius $\dmscr^{60,A}$ and the total charge radii $R_c$ are listed with the total uncertainties. Please note that there are diminutive corrections (flips in the last digit) in $\dmscr^{60,A}$ compared to \cite{kaufmann2020} caused by a correction in the analysis code.}
\begin{tabular}{l p{0.2cm} S[table-number-alignment = right] l p{0.1cm} S S}
\hline \hline
$A$ & & \multicolumn{2}{c}{$\delta \nu^{60,A}$/MHz} & & {$\dmscr^{60,A}$/fm$^2$} & {$R_c$/fm} \\
\hline 
58 & & -509.1(25) & \,\,\,\,\,[42] & & -0.275(8) & 3.770(2) \\
59 & & -214.3(27) & \,\,\,\,\,[22] & & -0.180(9) & 3.782(2) \\
60 & & 0.0 & & & 0.0 & 3.806(2) \\
61 & & 280.8(27) & \,\,\,\,\,[20] & & 0.082(5) & 3.817(2) \\
62 & & 503.9(25) & \,\,\,\,\,[39] & & 0.223(5) & 3.835(2) \\
63 & & 784.9(26) & \,\,\,\,\,[57] & & 0.277(8) & 3.842(2) \\
64 & & 1027.2(25) & \,\,\,\,\,[77] & & 0.367(10) & 3.854(2) \\
65 & & 1317.5(26) & \,\,\,\,\,[94] & & 0.385(18) & 3.856(3) \\
66 & & 1526.8(26) & \,\,\,\,\,[113] & & 0.493(17) & 3.870(3) \\
67 & & 1796.6(26) & \,\,\,\,\,[130] & & 0.514(25) & 3.873(3) \\
68 & & 1992.3(27) & \,\,\,\,\,[147] & & 0.619(24) & 3.886(3) \\
70 & & 2377.2(49) & \,\,\,\,\,[181] & & 0.806(24) & 3.910(3) \\
\hline
		\end{tabular}
	\label{tab:shifts}
\end{table}

{\it Theory. ---} 
\textit{Ab initio} approaches compute the mean-square charge radius $\mscr$ starting from the calculated point-proton mean-square radius $\langle r_\text{p}^2 \rangle$,
\begin{equation}
\mscr
=
\langle
r^2_{\text{p}}
\rangle
+
\langle
R^2_{\text{p}}
\rangle
+
\frac{N}{Z}
\langle
R^2_{\text{n}}
\rangle
+
\langle
r^2
\rangle_{\text{so}}
+
\frac{3 \hbar^2}{4 m_\text{p}^2 c^2} \; ,
\label{eq_mscr}
\end{equation}
where $\langle R^2_{\text{p}} \rangle$ and $\langle R^2_{\text{n}} \rangle$ are the mean-square charge radii of the proton and the neutron respectively, $\langle r^2\rangle_{\text{so}}$ denotes a spin-orbit correction~\cite{horowitz2012,hagen2015} and the last term corresponds to the relativistic Darwin-Foldy correction~\cite{Friar97}, with $m_\text{p}$ being the proton mass \cite{PhysRevLett.119.033001}.
The intrinsic (i.e. with respect to the center of mass) squared charge radius operator~\cite{hagen2010b,hagen2015,Cipollone15} is employed for $\langle r_\text{p}^2 \rangle$ in all calculations.
In the present work, the values of $\langle R^2_{\text{p}} \rangle = 0.709 \text{ fm}^2$~\cite{Pohl10, Xiong19} and  $\langle R^2_{\text{n}} \rangle = - 0.106 \text{ fm}^2$~\cite{Filin20} were used.

We employ the following two- plus three-nucleon (3N) interactions from chiral EFT:
(i) NNLO$_{\mathrm{sat}}$~\cite{ekstrom2015}, which gives a good description of charge radii in light and mid-mass isotopes but somewhat underbinds finite nuclei~\cite{hagen2015,GarciaCaChRadii2016,Morr17Tin,Soma20a,Soma20c,Heyl20Al,deGroote2020};
(ii) 1.8/2.0(EM)~\cite{Hebe11fits,Simo17SatFinNuc,Holt19drip}, and (iii) $NN$+$3N\text{(lnl)}$~\cite{Soma20a}, which reproduce ground-state and excitation energies throughout the medium- and heavy mass region, but generally underpredict absolute charge radii~\cite{Soma20c,Holt19drip,Miya21Heavy}.
The present work addresses a long sequence of charge radii along the Ni isotopic chain for the first time with three \textit{ab initio} techniques, using these three nuclear interactions. This provides a new, stringent accuracy benchmark of state-of-the-art methods which implement different computational schemes.
Importantly, a thorough evaluation of theoretical uncertainties is carried out for each many-body technique, as briefly described in the following.

The self-consistent Green's function (SCGF) approach~\cite{Dickhoff04, Carbone13, Soma20b} is a full-space correlation-expansion method applicable to the description of medium-mass nuclei~\cite{Raimondi19a,Barbieri19,Idini19,Chen19,Sun20,Soma20a,Soma20c}.
Nickel isotopes were recently addressed in Ref.~\cite{Soma20a} where the calculation of radii, however, was not optimized and lacked theoretical uncertainties.
Here, we present 
nickel charge radii with a full analysis of basis convergence and an assessment of associated theoretical errors.
To this end, SCGF calculations are performed in the Gorkov ADC(2) scheme~\cite{Soma11, Soma14a} using a spherical harmonic-oscillator basis including up to 14 major shells ($e_\text{max} \equiv \text{max}(2n + l) =13$), with matrix elements of three-body operators further restricted to $e_\text{3max}=16$.
Theoretical errors comprise uncertainties arising from both many-body and model-space truncations.
The former are estimated from differences between ADC(2) and ADC(3)~\cite{BarbCarbLNP, Raimondi18} calculations, available for closed-shell isotopes.
The latter are evaluated from a range of oscillator frequencies, $\hbar\Omega$, within 2~MeV from the optimal values.

The valence-space in-medium similarity renormalization group (VS-IMSRG) method~\cite{Tsuk12SM,Bogn14SM,Stro16TNO,Herg16PR,Stro17ENO,Stro19ARNPS,Miya20MS} decouples a valence-space Hamiltonian and consistent operators from the full-space problem via an approximate unitary transformation. 
To obtain charge radii, we first decouple the core and valence-space intrinsic proton mean-squared radius operator and then apply Eq.~\eqref{eq_mscr}.
We use the IMSRG(2) approximation where induced operators are truncated at the two-body level and the ensemble normal ordering procedure~\cite{Morr15Magnus,Stro17ENO}, which captures the physics of 3N forces between valence particles. We take the neutron $p_{3/2}$,\,$p_{1/2}$,\,$f_{5/2}$,\,$g_{9/2}$ valence space outside a $^{56}$Ni core, decouple a valence-space Hamiltonian for each isotope studied and diagonalize with the {\sc kshell} code~\cite{Shimizu2019} to obtain
expectation values for the intrinsic proton mean-square radius operator. While model-space uncertainties are obtained analogously to SCGF, errors due to the many-body method cannot be estimated currently \cite{Heinz:2021xir}. The $e_\text{max}$/$e_\text{3max}$ and basis choices are as in Ref. \cite{Holt19drip}.

The coupled-cluster method performs a similarity transformation of the Hamiltonian and decouples a 
reference state from its $n$-particle--$n$-hole ($n$p-$n$h) excitations~\cite{coester1958,kuemmel1978,bartlett2007,hagen2014}. 
This method was used to compute the structure of doubly-magic nuclei and their neighbours \cite{hagen2012b,hagen2016b,Morr17Tin,gysbers2019,payne2019,kaufmann2020} and can also be extended to open-shell nuclei~\cite{novario2020}.
Our calculations for nickel isotopes employ
a single-particle basis of up to 13 harmonic oscillator shells with a frequency $\hbar\Omega=16$~MeV; matrix elements of three-nucleon forces are truncated at $e_{3\rm max}=16$. We start from an axially symmetric Hartree-Fock reference, normal-order the resulting Hamiltonian with respect to this state, and truncate it at the two-body level~\cite{hagen2007a,roth2012}. The ensuing coupled-cluster calculations employ the CCSD approximation, i.e., 1p-1h and 2p-2h excitations of the reference are fully decoupled. While this captures (only) about 90\% of the correlation energy, the omission of 3p-3h excitations has a much smaller effect on radii and introduces an estimated 1\% uncertainty. Uncertainties from the finite model space are estimated from the difference between calculations in 11 and 13 harmonic oscillator shells. Overall, we estimate coupled-cluster uncertainties on $R_c$ to be +2\%/-1\%.

The fourth theory considered is nuclear DFT
\cite{Bender2003}.
Here, we focus on non-relativistic energy density functionals (EDF)  and employ two
EDF parametrizations, namely SV-min \cite{Kluepfel2009} as representative of the
widely used Skyrme functionals and Fy($\Delta r$, HFB) as the recent
example of a Fayans functional \cite{Miller2019}.  Both have the basic
structure in common and
are calibrated with the same fitting
strategy to the same large body of nuclear ground state data (energy,
radii, surface thickness, ...) as described in
\cite{Kluepfel2009}. The Fayans functional Fy($\Delta r$, HFB) differs
in that it contains additional gradient terms in surface and pairing
energies \cite{Fayans1998,Fayans2000} and that isotopic shifts of
charge radii in the calcium chain were added to the optimization data set. 
%
The rms charge radii are computed directly from the nuclear charge form factor. The latter is obtained from folding the proton and neutron densities with the intrinsic charge and magnetic densities of the nucleons, for details see Ref.\ 
\cite{dftRc_2021}. The calculations are done with codes allowing for deformed ground
states, for SV-min with SkyAx \cite{Reinhard2021}, and for Fy($\Delta r$, HFB) with a version of HFBTHO \cite{HFBTHO} extended to Fayans
EDF. 
Results for spherical nuclei have been
counter-checked with our spherical BCS/HFB code, the one which was
used for the calibration of both functionals
\cite{Kluepfel2009,Miller2019}.
%
DFT parametrizations carry statistical uncertainties \cite{Kluepfel2009}
as well as systematic errors
related to principle limitations of the model \cite{PhysRevC.88.031305}. 

\begin{figure}[t]
	\centering
		\includegraphics[width=\columnwidth]{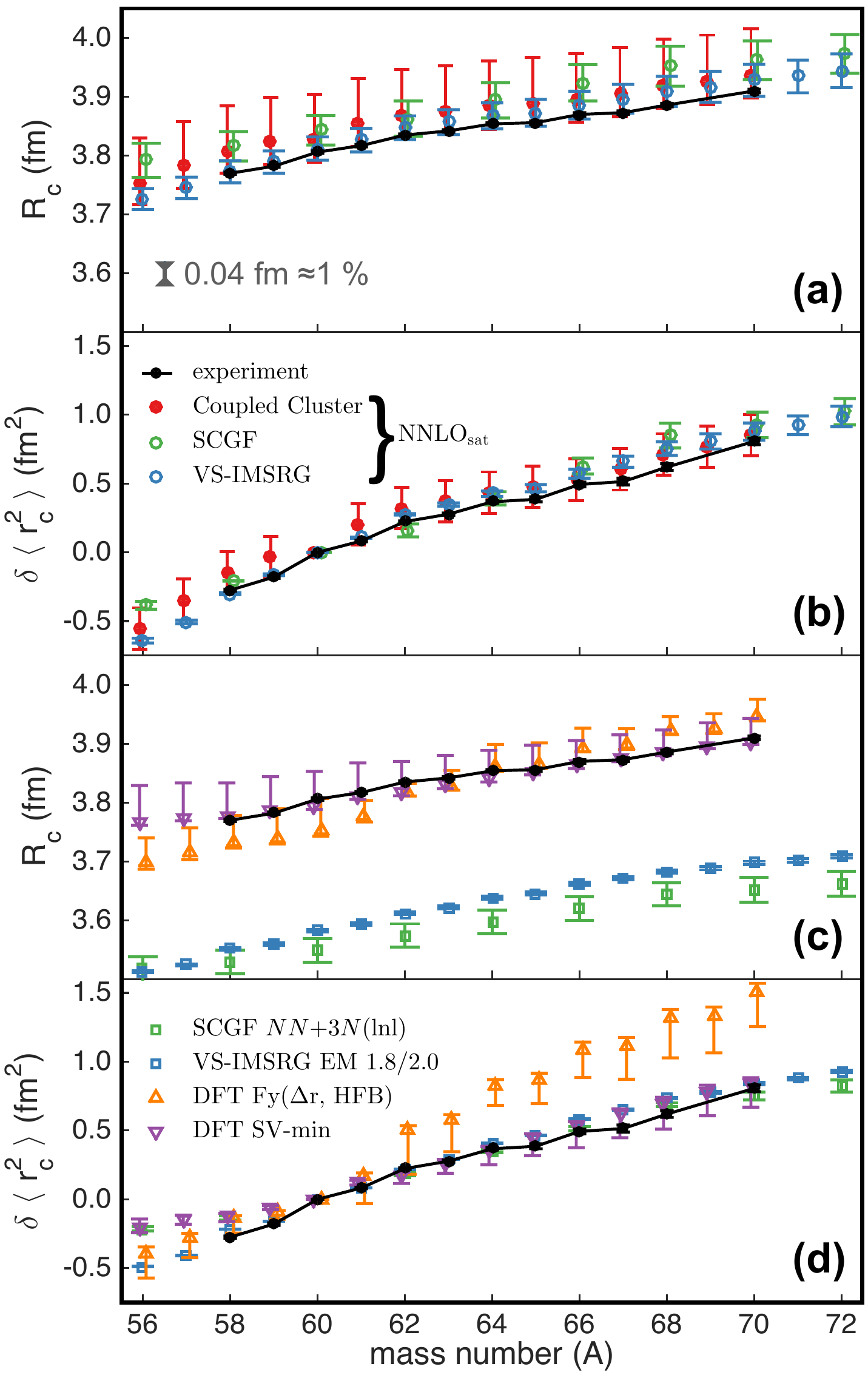}
	\caption{Nuclear charge radii $R_c$ and differentials $\dmscr^{60,A}$ of Ni isotopes with respect to $^{60}$Ni as reference. Experimental data are compared to theoretical results. 
	See text for details.}
	\label{fig:niChargeRadii}
\end{figure}

{\it Discussion. ---} 
Theoretical and experimental nuclear charge radii are compared in Fig.\,\ref{fig:niChargeRadii}. Charge radii $R_c$ provide a comparison on the absolute scale, while the differential charge radii $\dmscr$ probe local variations in the nuclear charge distribution more closely, since various theoretical uncertainties cancel in $\dmscr$. For instance, the errors on $R_c$ in DFT contain a sizable, nearly constant offset along the chain reflecting a certain vibrational softness
for all Ni isotopes. These vibrational corrections enhance total radii and are thus predominantly positive, but are greatly reduced in $\dmscr$.

For SV-min based DFT as well as for all \textit{ab initio} calculations based on NNLO$_{\mathrm{sat}}$, the overall agreement with experiment is very good. For both, $R_c$ and $\dmscr$, the experimental values are within, or very close to, the theoretical error band, which is of the order of $\approx1$~\%. The same holds for the differential radii $\dmscr$ when considering \textit{ab initio} results for the other employed nuclear interactions, see Fig.\,\ref{fig:niChargeRadii}(d),
while those deviate notably from experiment in the absolute charge radii $R_c$, as shown in Fig.\,\ref{fig:niChargeRadii}(c). This is in line with the expectation from previous work \cite{Soma20c,deGroote2020}.

Within the same nuclear many-body method, calculations of $R_c$ with NNLO$_{\mathrm{sat}}$ disagree with the results of both 1.8/2.0(EM) and $NN$+$3N\text{(lnl)}$. This illustrates the sensitivity of $R_c$ on the accurate encoding of the relevant physics for medium-mass nuclei into nuclear forces~\cite{ekstrom2015,Simo17SatFinNuc}.
On the other hand, a comprehensive assessment of  uncertainties due to a many-body method itself remains a challenge. 
Employing the same nuclear interaction in conjunction with different many-body methods is one way to evaluate many-body uncertainties. 
As shown in Fig.\,\ref{fig:niChargeRadii}(a) and \ref{fig:niChargeRadii}(b), the results of 
SCGF, VS-IMSRG, and coupled cluster theory, all utilising NNLO$_{\mathrm{sat}}$, agree with each other within the theoretical uncertainties, thus, providing strong evidence for the accuracy of the  methods. Small differences can be seen for $^{56}$Ni where uncertainties of SCGF and VS-IMSRG do not overlap. Note that the error bars in VS-IMSRG account for model-space uncertainties only. We have confirmed that the latter are consistent in size across different methods.

With respect to nuclear charge radii, the Fayans functional has been very successful in describing an odd-even staggering as well as characteristic kinks typically found at shell closures \cite{GarciaCaChRadii2016,Miller2019,deGroote2020,PhysRevLett.121.102501,PhysRevLett.122.192502}. In contrast, DFT utilizing Skyrme functionals such as SV-min generally fails to reproduce both. However,  compared to the large odd-even staggering in Ca \cite{GarciaCaChRadii2016,Miller2019} or the sizeable kink at $N= 82$ in Sn \cite{PhysRevLett.122.192502}, charge radii along the measured Ni isotopes do not exhibit these features very prominently. Interestingly, the SV-min follows in this case the experimental trend more closely compared to Fy($\Delta r$, HFB), see Fig.\,\ref{fig:niChargeRadii}(d). On closer inspection, analogous conclusions for the mid-shell region also hold for the charge radii of Cu \cite{deGroote2020} and  Sn isotopes \cite{PhysRevLett.122.192502}. A potential  deficiency of the present Fayans functional could be its lack of an isovector component in its pairing part \cite{ReinhardNazarewicz2017}. 
Hence, future efforts in Fayans-based DFT will focus on pinning down the (presently unused) isovector term in the pairing functional, see Ref. \cite{Bollapragada_2021}.

{\it Summary ---} 
Collinear laser spectroscopy of short-lived nickel isotopes $^{58-68,70}$Ni was performed.
The extracted nuclear mean-square charge radii $R_c$ 
benchmark theoretical work applying density functional theory as well as three \textit{ab initio} methods.
When the same chiral EFT-based nuclear potential NNLO$_{\mathrm{sat}}$ is utilized in all \textit{ab initio} calculations, their results show excellent consistency and they agree well with experiment. Calculations exploiting other nuclear interactions perform equally well for $\dmscr$, 
but struggle in reproducing the absolute radii. Interestingly, in the absence of prominent features such as unusually-large odd-even staggering or kinks in $R_c$, which have been successfully described by Fayans-based functionals, Skyrme-based DFT yields results closer to experiment. Overall, this comparative work combining experiment, density functional theory and \textit{ab initio} calculations establishes a theoretical accuracy of $\sim1$\% for the description of nuclear charge radii in the Ni region. 
\section{Acknowledgments}

We acknowledge the support of the ISOLDE Collaboration and technical teams, the ISOLTRAP group, and funding from the European Union's Horizon 2020 programme under grant agreement no.\ 654002. We thank the Physikalisch Technische Bundesanstalt (PTB) Braunschweig for the loan of a precision high-voltage divider. This work was supported by the Max-Planck Society, the Deutsche Forschungsgemeinschaft (DFG, German Research Foundation) -- Project-Id 279384907 -- SFB 1245, the Collaborative Research Center [The Low-Energy Frontier of the Standard Model (SFB 1044)], the Cluster of Excellence ``Precision Physics, Fundamental Interactions, and Structure of Matter'' (PRISMA$^+$ EXC 2118/1) funded by DFG within the German Excellence Strategy -- Projektnummer 39083149 --, the BMBF under Contracts No.~05P18RDCIA, 05P18RDFN1, and 05P19RDFN1, the FWO (Belgium), GOA 15/010 from KU Leuven, NSERC, and the Office of Nuclear Physics, U.S.~Department of Energy, under
grant  Nos.  DE-FG02-96ER40963, DE-SC0013365, No. DE-SC0018083, and DESC0018223 (NUCLEI SciDAC-4 collaboration). Computer time was provided by the Innovative and Novel Computational Impact on Theory and  Experiment  (INCITE)  programme.   This  research used resources of the Oak Ridge Leadership Computing Facility located at Oak Ridge National Laboratory, which is supported by the Office of Science of the Department of Energy under contract No.  DE-AC05-00OR22725. The calculations presented in this work were also performed on ``Mogon II'' at Johannes Gutenberg-Universit\"{a}t in Mainz. 
Computational resources for DFT calculations were partly provided by the CSC-IT Center for Science Ltd. (Finland). SCGF calculations were performed by using HPC resources from GENCI-TGCC, France (Contract No. A007057392) and at the DiRAC Complexity system at the University of Leicester, UK (BIS National E-infrastructure capital grant No. ST/K000373/1 and STFC grant No. ST/K0003259/1).
This work was also supported by consolidated grants from STFC (UK) - ST/L005516/1, ST/L005670/1, ST/L005794/1, ST/P004423/1, and ST/P004598/1.
TRIUMF receives federal funding via a contribution agreement with the National Research Council of Canada.

\bibliographystyle{apsrev4-1}
\bibliography{NickelRadii,master,stephanBiblio}

\end{document}